\documentclass{article}

\usepackage{graphics}
\graphicspath{ {images/} }
\usepackage{float} 
\usepackage{multirow} 
\usepackage{longtable}
\usepackage{adjustbox}

\usepackage{natbib}

\usepackage{setspace}
\usepackage{geometry}
\usepackage{afterpage}

\usepackage[T1]{fontenc}
\usepackage{textcomp}

\usepackage{pifont}

\usepackage{algorithm}
\usepackage{algorithmic}
\usepackage{amsmath}
\usepackage{amsfonts}
\usepackage{amssymb}

\usepackage{enumitem}
\usepackage{url}
\usepackage{rotating}

\usepackage{caption}
\usepackage[english]{babel}
\usepackage{array}

\usepackage{subcaption} 
\usepackage[colorlinks=TRUE, linkcolor=blue, urlcolor=blue, citecolor=blue]{hyperref}

\usepackage{footnote}
\makeatletter

\normalsize 
\usepackage{titling,lipsum}
\makeatother
\begin{document}

\title{The more you ask, the less you get: the negative impact of collaborative overload on performance}
\author{
\large
\textsc{Anna Velyka $^{1,3}$ and Marco Guerzoni $^{1,2}$}\\ 
\\[2mm] 
\normalsize {$^{1}$ Despina, Big Data Lab, Department of Economics and Statistics "Cognetti de Martiis"} \\
  \normalsize {$^{2}$ ICRIOS, Bocconi Univerisity}\\
 \normalsize {$^{3}$ aizoOn Technology Consulting} \\ 
\vspace{-4mm}
}

\date{}

\begin{titlingpage}
\maketitle
\vspace{2cm}
\end{titlingpage}
\pagebreak
\begin{abstract}
\noindent 
\par This paper is about the possible negative impact of excessive collaboration on the performance of top employees. With the rise of participatory culture and developments in communications technology, management practices require greater conceptual awareness about possible outcomes of increased organizational interconnectivity. While there exists a sound theoretical basis for possible burdens brought by collaborative overload, the literature never really manage to measure and empirically test this phenomenon. We address this gap by developing a methodological framework for the identification of organizational actors at risk of operational capacity overload. Drawing on social network analysis as the widely applied approach for the estimation of employees' involvement in the information exchange networks, this paper describes potential personal and organizational causes leading to the emergence of collaborative overload. Relying on primary data gathered through a survey conducted among employees in a large insurance company, we present a  testable model for overload detection. A second merit of the paper consists in finding a novel identification strategy for empirical work on cross-sectional network data, which often face the issue of endogeneity.
\par This research suggests that active collaborative activity does not cause a decrease throughout every aspect of performance. We found that expertise sharing depends on a few key players who take core knowledge assets upon themselves and thus run higher risks of exposure to overload.  performance. 
\\[4mm] 
\end{abstract}

\pagebreak
\section{Introduction}

\par There is an increasing awareness of the strategic role organizational communication networks play in the process of knowledge diffusion. 
Formal organizational structure establishes the order for interpersonal communication in the workplace to facilitate knowledge sharing and defines the right people to execute appropriate steps in managing knowledge-intensive processes \citep{DiCiccio2012}. 
However, many organizations experience inconsistency between the formal and informal structure of interactions that lead to the emergence of a limited number of "go-to" employees who act as a reference point for expertise in a particular area \citep{Oldroyd2012, Cross2013}. These knowledge workers are in high demand among colleagues, and they dedicate a significant amount of time processing external requests for expertise sharing. 

\par Rewards and recognition program defines compensation mechanisms for an acknowledgment of workers' achievements and realization of optimal performance \citep{Milne2001}. The majority of such mechanisms fail to account for the expertise sharing activity and thus fail to reveal the full corporate value of knowledge workers. The difficulty of evaluating and managing something that can not be measured complicates the inclusion of the relationship-building competency in rewards and recognition programs  \citep{Milne2001}. Social network analysis provided the means to overcome a methodological challenge presented by the intangible nature and subjectivity of knowledge transfer via personal interactions and quantify its potential contribution.

\par While the analysis of interpersonal communication networks acquires greater importance in the evaluation of organizational performance,  scientific community is warning that knowledge sharing through interpersonal communication in the workplace may come at a cost. Extensive collaboration carries the risk of negative externalities resulting in unequal resource distribution, collective rivalry \citep{Adler2002}, and decline in performance for those employees who experience social capital overload \citep{ Cross2013, Oldroyd2012}. The concept of overload appears multifaceted in academic literature. Different authors invoke it as an umbrella term accounting for a decrease in information processing activity \citep{Eppler2004}, intra-role conflict \citep{Jones2007}, and excessive collaboration demand \citep{Cross2013}. Whereas recent research in the management science domain present an extensive overview of potential factors leading to overload \citep{Allen2003, Eppler2004, Jackson2012}, there is a lack of empirical research that produce evidence of overload existence and measures the effect size of these factors on individual and organizational performance, thereby assisting managers and corporate leaders in creating a productive knowledge-driven collaborative environment. 

\par The present paper calls into question the notion of collaborative overload, which occurs when worker's performance drops as the result of increasing social capital expressed through the growing number of incoming requests from the colleagues for expertise sharing. We define performance as the ability of a worker to satisfy incoming demand for the information. Our knowledge of the negative effect of collaboration on performance is largely based on very limited empirical evidence, which casts doubt on the validity of previously developed theoretical assumptions. The aim of the research is thus to contribute to the evidence-based analysis of intraorganizational collaboration aspects that may lead to a decline in individual performance. This paper invokes the structural dimension of social capital to reveal the pattern of knowledge transfers between workers \citep{Burt1992, Nahapiet1998}. To analyze who, how, and for which information workers reach out to, we collect the data using a survey of 303 employees in a big Italian insurance company. We present an ordinal logistic regression model to estimate the relationship between the increase in social capital and the efficiency of knowledge transfers, controlling for individual characteristics of workers and task diversity.

\par Our research supports the theoretical assumption that the social embeddedness of workers in organizational knowledge transfers does not follow a normal distribution, which results in the emergence of a few key expertise holders. We find evidence of the collaborative overload for those workers whose growing engagement in knowledge sharing activities leads to a decrease in their performance.
This study extends existing theory about the risks expected from solidarity benefit of social capital \citep{Adler2002, Pillai2017}. The present paper introduced quantitative methods to capture various organizational and personal factors affecting negatively individual processing capacity 
and reveal particular aspects of performance that are the most vulnerable to overload. 

\par Understanding the effect of collaboration on employees' performance is of great importance for effective human and knowledge resource management. We contribute to the existing literature on the topic by providing the empirical analysis of communication patterns in the knowledge-intensive industry, identifying aspects potentially vulnerable to collaboration overload. Moreover, in the process of doing so, we develop an identification strategy, which has a broader scope than this paper. \citet{carpenter2012social} creates a taxonomy of research in a social network, according to which this paper will be defined as social capital research with a focus on interpersonal relationships. In this frame of research, \citet{carpenter2012social} highlights that a key empirical issue is a likely endogeneity stemming from the simultaneity of performance indicators and network constructs, and they suggest the use of an instrumental variable approach as it is typical in management science \citep{bascle2008controlling, shaver1998accounting}. Nevertheless, they acknowledge that there are not many examples of instrumental variables application, being a notable exception \citet{zaheer2009network}, which exploits the lagged structure of the data as an instrument. In the methodological section, we put forward an identification strategy that relies on the concept of homophily to both derive a suitable instrument for the matter of this paper, and provide a possible wider application in network studies applied to organizational science.

\par In the next paragraph, we sketch the theoretical framework and, thereafter, we present the data and the empirical strategy. Results and conclusions follow.

\pagebreak

\section{Theoretical Framework}
\par The debate between communication effectiveness and organizational performance dates back to \citet{Downs1988, Redding1988}, while before scholars neglected the relation between the individual behavior and the whole organization. Furthermore, the difficulty in defining communication's contribution to the overall performance, owed to an elusive nature of social contact, hindered attempts in quantifying this relationship.

\par However in the present time, encouraging teamwork and peer support is considered a precondition for establishing effective organizational culture \citep{Tjosvold1989}. Numerous publications praised management intervention in intraorganizational collaborative practices, anticipating positive externalities for knowledge transfers, employees' personal development, and the diffusion of innovation \citep{Goldhaber1988, Downs1988, Tjosvold1989, Dooley2002}.  

\par More recently, the notion of social capital has been eventually allowing to study the value generated by collaboration among employees and overcome also the boundaries for a quantitative analysis of the process of employees' interactions \citep{ Abbasi2014}. Previous research described the applicability of employees relationship graphs as an accurate technique to reveal the discrete nature of organizational interactions \citep{ Freeman1978, Ibarra1993, Toni2010, Oldroyd2012,  Abbasi2014, Yamkovenko2015}. By unveiling seemingly elusive personal networks, sociologists established the crucial role of collaborations for knowledge and information exchange \citep{Granovetter1983}; described the importance of brokerage role for access and control of novel information \citep{Burt1992} and traced an eventual innovation enhancement \citep{Tsai1998}. More recent works made a step forward and shifted the focus on the explanation of network formation and the outcome that network structure may have on organizational performance \citep{Monge2003}.

\par The network of impersonal linkages between units reveals both the composition and the patterns of connections and it forms the structural dimension of social capital \citep{Nahapiet1998}, which ensures exchanging and transferring knowledge resources \citep{Ansari2012} and facilitates access to colleagues who poses desired expertise \citep{Andrews2010}. 
The relational dimension of social capital retains information about the qualitative aspect crucial for the emergence of social cooperation. It attributes the features of interpersonal relationships and accounts for normative behavior based on trust, obligations, and expectations which is overlooked by the structural dimension \citep{Lee2008}. The analysis of this dimension reveals the willingness of an individual to prioritize the requests of the colleagues while setting aside individual tasks and thus favoring collective goals \citep{Lazarova2009}.

The evolution of social structures in the organizational setting may lead to the establishment of an efficient performance pattern and contribute to the development of a highly resistant system \citep{Fukuyama2002}. Thus, social capital is prone to produce both positive and negative externalities on different levels relating to individuals, groups, and organization performance \citep{Coleman1988}. While the positive externalities of social capital management are well discussed, academic and business literature often overlooks the other side of the coin \citep{Adler2002}. The risks connected to social capital arise as a result of information, power or solidarity benefits. This paper calls into question contingencies arising from solidarity benefits resulting in over-embeddedness, dominance, and exclusion of employees, finally leading to the exhaustion of a collaboration.

\par One of the pioneering suggestions about the negative externalities of excessive communication originates in the early work by \citet{Wiio1978}, where the author invoked information processing capacity limit and questioned the universality of communication increase panacea.  Since individuals possess limited cognitive capacity, time, and knowledge resources, thy are encouraged to seek expertise from their colleagues to optimize the time required for a decision making and the quality of the decision itself \citep{Borgatti2003}. The literature offers a myriad of conflicting terms to define a burden that results from this intensified communication between employees. While we do not intend to discard previous definitions, we stress the importance of an accurate theoretical rationale in defining one. Though the terms information-, collaboration-, and role-overload overlap, they originate from different analytical facets, and a sound conceptualization of the term restricts it to a specific context, allowing for a careful selection of compatible variables for the analysis and limits the scope of possible practical implications.

\par The most commonly applied umbrella term to describe the conflict between individual processing capacities and processing requirements is information overload \citep{Eppler2004}.  It originates from the inverted u-curve dependence between effective decision-making and information exposure. An upward trend is expected to last up until the point when the further delivery of new information fails to be integrated, and the performance of the actor starts to decline. \citet{Eppler2004} concluded that the overload results from the decrease in processing capacity of the receiver, his ability to integrate new information and to properly allocate time for processing.
\par In the organizational setting employees may be exposed to various information flows, originating from numerous sources, transferred through different communication channels. When evaluating employees professional communication, we fail to account for the information the actor is consuming for personal interest. Such research would require a more detailed type of questionnaire or extensive observation records that are difficult to implement in the organizational setting without incurring losses in quality of results.  This research opts for the term \textsl{collaborative overload} due to the inability to fully account for the information flows employees may experience on a daily basis. 

\par This paper seeks to address the setting strictly linked to the \textsl{professional service industry} that according to \citet{Oldroyd2012} is most likely to produce a limited number of key expertise holders.  The object of analysis originates from relational data of the organizational social system that reveals the informal structure of communication taking place between employees. The expectancy-value theory treats individuals as goal-oriented and defines two appraisals guiding their behavior: a belief that action will result in a conceived outcome and the evaluation of possible positive or negative degree of this outcome \citep{Palmgreen1984}. Thus, we assume that advice-seekers will be granting preferential treatment to those colleagues who are expected to deliver a more significant probability of a successful outcome. Each employee trying to maximize individual performance will tend to favor those channels that bring the highest quality of collaboration, thus contributing to the disproportionate distribution of interactions. Some actors will be involved in the knowledge sharing more than others and therefore, restrict the decision-making process and concentrate essential knowledge resources in the minds of a few. The pressure of collaborative demand towards a limited group of individuals will be growing, requiring from them a higher level of information processing capacity, time management, and balance of focus. \citet{Cross2013} introduced a term collaborative overload to refer to possible unintended outcomes of increasing collaboration demand on high performing employees.

Previous works have primarily contributed to the theoretical analysis and conceptual discussion but failed to provide a reliable empirical prove to understand circumstances under which productivity of focal actors decrease \citep{Eppler2004}. Several studies proposed testable models for the evaluation of possible effects of social capital abundance on employees performance \citep{Speier1999, Allen2003, Eppler2004, Oldroyd2012}. Collaboration in the knowledge-intensive organizations is expected to have a cross-level pattern, neglecting hierarchical structure \citep{Agneessens2012} and social network analysis can reveals these unobserved patterns of informal interactions among employees \citep{Toni2010}:  Focal actors, informally recognized as such by other colleagues, receive an abundant number of incoming information requests, which contribute to expertise accumulation and thus positively affecting the chances to be approached again increasing their visibility in the organizational network. Eventually, a growing frequency of incoming requests from colleagues conditions individual processing capacity of the top performers \citep{Eppler2004}. In their investigation into the social capital, \citet*{Oldroyd2012} described how a virtuous cycle of employees stardom might turn into a vicious cycle of overload. In the remaining of the paper we present data and develop a testable model to show that focal actors in the informal network of information exchange do exhibit an above-average performance, which is however doomed to decrease when a threshold of collaborative overload is reached.

\pagebreak
\section{Data and Methodology}
To test the hypothesis about the possible negative externalities of intensive collaboration, we collected data about organizational communications in a large Italy-based international insurance company. This professional service industry relies heavily on knowledge as an input and as the final output with the goal of customized solutions' delivery \citep{Empson2001}. Organizations producing tangible products (labor and capital-intensive industries) generate innovations by selecting the most lucrative options through internal R\&D activities \citep{Muller2009}, while knowledge intensive business services prioritize organizational innovations that come as the result of a unique combination of technology and human skills \citep{Muller2009}. Thus, a professional service industry provides the setting where human capital is a dominant factor responsible for the generation of core knowledge assets \citep{Oldroyd2012} making it a reasonable test-bed for the exercise carried out in the paper.

The evaluation of knowledge exchange through the prism of social network analysis has been widely used to map informal networks \citep{Granovetter1983, Krachardt1993, Abbasi2014}, to identify influential actors involved in decision-making activities \citep{Toni2010}, to capture information exchange \citep{Oldroyd2012}, and to evaluate the organization's inclusion, connectedness and diversity \citep{Yamkovenko2015}. We define collaboration as the process in which organizational members initiate relational ties with colleagues in search of information. Therefore, the basic unit of analysis in this study is intraorganizational communications for expertise sharing. 

\par We made use of two sources of data. The primary data come from a survey providing information on informal interaction, task constraints, and a self-assessed quality evaluation on interaction. We merged this data with the second source which summarizes personal characteristics of the actors. As for the survey, we used purposive sampling to identify the organizational actors of interest in our study. To identify key knowledge holders engaged in sharing professional expertise from the total population of employees we selected 303 professionals dealing with insurance claims settlement. Selected participants were asked to cite not more than five colleagues for each of the four major professional areas of the company. They were asked to concentrate primarily on the interactions and exchanges of information that happen beside the formal meetings, training courses and practices determined by the organizational rules \citep{Krachardt1993, Cross2002, Kase2009, Toni2010, Oldroyd2012, Behrendt2014, Yamkovenko2015}. By asking participants to focus on informal connections we recreated the exchange of the information not registered daily or established in the \textbf{ad hoc} documentation. The response rate was 79.6\%, corresponding to 241 individuals generating 1437 citations.

\par Specifically, respondents were asked to mention 5 colleagues to whom they came for knowledge and information required to perform a certain task in the previous six months. For each person mentioned, respondents were asked to evaluate 4 items\footnote{see Appendix \ref{survey} for a detailed description of the survey}:
\begin{enumerate}
    \item The frequency of interaction on a scale from 1 to 4;
    \item The satisfaction of the answers received on a scale from 1 to 6 \footnote{We chose an even scale of evaluation to avoid neutral responses as a way to skip the question.};
    \item The type of information delivery requested in the question. The possible answer range from a very simple type of knowledge transfer such as "Simple instruction" up to a complex involvement of the co-worker such as "Step-by-step follow-up of the specific activity and verifying results" or "Practical demonstration with exercises and experiments on cases";
    \item The communication channel, covering the usual mode of communications (email, telephone, personal meeting); 
\end{enumerate}

\par Item 1, the frequency of collaboration between two individuals allows us to create a proxy for the burden of communication carried by different workers. One of the main factors in rising demand for an employee's processing capacity is the quantity and the frequency of the interactions required to satisfy the request for expertise sharing. We created the variable $Load$, which accounts for the intensity of collaborations experienced by the employee. Overload is expected to appear as the number of concurrent tasks and interruptions is rising and reducing the concentration of an employee, who experiences repeated context switching. \citep{Eppler2004, Cross2016}.

\par Item 2 attempts to measure the worker's performance.  We opted for a subjective measurement of employees productivity through an individual opinion about request satisfaction. Thus, the variable of interest, $Efficiency$, measures the average efficiency of requests to collaborate as subjectively measured by the respondent of the survey.

\par Item 3 controls for the complexity of the knowledge involved in the process. \citet*{ Schneider1987} argued that the requirements for the information processing capacity are growing proportionally with the growth of the amount and the nature of information characterized by novelty, uncertainty, and complexity. Due to the higher requirements that these factors pose on individual cognitive capacity and time, they might also increase the possibility of overload occurrence. Due to statistical constraints explained in the next paragraph, we created the variable $Knowledge$ scaling down the initial six categories into two. The first category combines simple, one-time and more detailed instructions and explanations, to reflect information based on recurrent organizational practices. The second category, bringing together practical demonstration, analysis of the specific problem and step-by-step follow-up, accounts for more complex interactions with higher requirements to respondents personal resource.

\par We used item 4 to create the variable $Channel$ to control for the  communication technology utilized to transmit information since it can create an additional constraint increasing the probability of overload occurrence \citep{Eppler2004}. \citep{Allen2003} anticipated potential risk of overload, due to the inefficient and unproductive use of information technology. By asking participants to indicate the means of communication used we determined the frequency and time required for interaction.

\par Finally, organizational status is considered a powerful predictor of the strength of an employee's position \citep{Krackhardt1990}. In their research on intraorganizational advice relations, \citet* {Agneessens2012} assumed that the status and its strength could be defined by characteristics, leading to its emergence. The interest of this research is the social status, defined by competences and expertise, gained through the organizational learning process. Based on Berger's theory of expectation states, \citet* {Agneessens2012} argued that regardless of being subjectively defined, a status can be derived from the general consensus. Employees' social worth and expectations are modeled through such variables as age and gender \citep{Downs1988}. Thus, as controls, we included  $Seniority$, $Age$, and $Gender$ as individual characteristics that indicate informal leaders, social value and that might partially explain the centrality in the organizational network and the performance of the agents. There are other variable, that we use to build the instrument in the regression analysis. $Location$ describe the city where the employee works, $Management$ is dummy variable with value 1 for high level management, $Level$ indicates the pay grade level which corresponds with the rank in the hierarchical structure of the company.

\subsection{Descriptive statistics}
\begin{figure}[H]
\caption{Distribution densities of Age and Seniority for female and male participants.}
\label{fig:age.sen}
\includegraphics[width=0.6\textwidth]{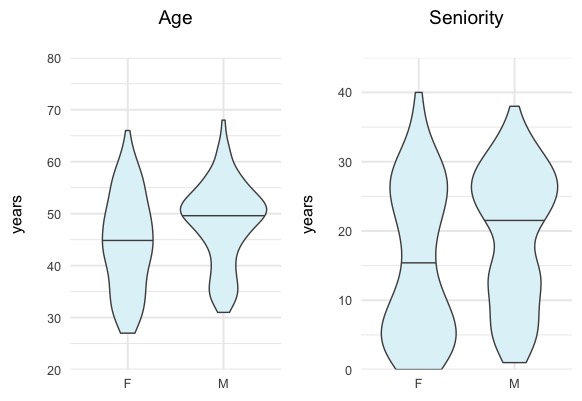}
\centering
\end{figure}

The  sample consisted of 303 specialists engaged in the four principal areas of the company's expertise (anti-fraud, reserve management, civil liability insurance, property damages liquidation). 60\% of the respondents were male. The age of the sample varied from 27 to 68 years, with a mean of 46 years. 35\% of the workers had less than 10 years of working experience in the company, and 11\% possessed more than 30 years. The majority of women have lower organizational tenure: 50\% of females had less than 15 years of working experience inside the company, compared to 22 years for male (See Figure \ref{fig:age.sen}). 
 
\par
\begin{figure}[h]
\caption{Network of collaboration and Efficiency}
\label{fig:network}
\includegraphics[width=0.6\textwidth]{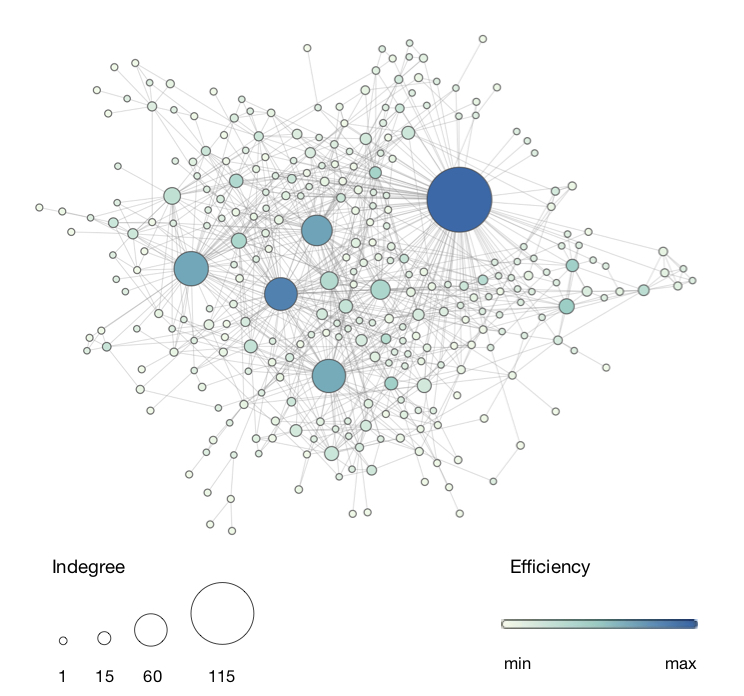}
\centering
\end{figure}

 As result of the survey, we built a network of employees' interactions, accounting for 1437 directed edges. Figure \ref{fig:network} shows the network of informal communication based on the survey. The size of the nodes describes the number of contact requests and the color the average efficiency of interaction. As in most real networks, the degree distribution is highly right-skewed. The majority of agents in the organization network have low degrees while a small but significant fraction have an extraordinarily high degree.
The figure clearly shows that the distribution of the collaborative burden is mainly on a few players. As expected, those key players exhibit a larger efficiency, since it is fairly reasonable to assume that they became central in the network precisely because of their reliability.

\par To account for collaboration activity, for each node we calculated the variable $Load$ by summing the frequency of each incoming edges. This variable was unequally distributed over the sample of employees, suggesting the presence of loaded top performers in the organization. 

\begin{figure}[H]
\centering
\begin{minipage}{.5\textwidth}
  \centering
  \includegraphics[width=.8\linewidth]{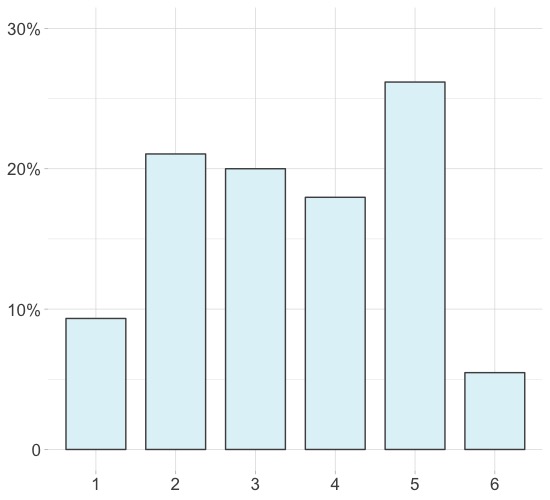}
  \captionof{figure}{Knowledge complexity}
  \label{fig:info}
\end{minipage}%
\begin{minipage}{.5\textwidth}
  \centering
  \includegraphics[width=.8\linewidth]{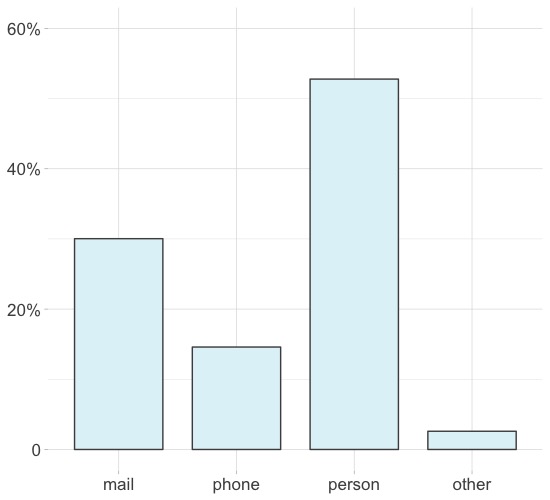}
  \captionof{figure}{Mode of communication}
  \label{fig:channel}
\end{minipage}
\end{figure}

\par Figure \ref{fig:info} depicts the distribution of the categorical variable  $Knowledge$. Simple instructions (category 1) and step-by-step follow-up (category 6) were cited by less than 10\% of respondents each. The most popular category 5 - analysis of the specific problem and definition of how to deal with it - resulted in 30\% of citations. Half of the requests concentrate on the first three categories dealing with simple instructions, information which should be codified and easily retrievable through organizational wikis. 
\par Figure \ref{fig:channel} depicts the type of communication channel used. Half of respondents indicated a preference for personal meetings as the way to contact colleagues, while the usage of email (30\%) prevailed over the telephone (15\% ). 

\par Tables \ref{tab: nominal} and \ref{tab: cont} summarizes the descriptive statistics for the variables. 

\begin{table}[H]
\centering
\caption{Descriptive statistics: categorical variables.}
\begingroup\footnotesize
\begin{tabular}{llrrr}
 \textbf{Variable} & \textbf{Levels} & $\mathbf{n}$ & $\mathbf{\%}$ & $\mathbf{\sum \%}$ \\ 
  \hline
Efficiency & 1 & 35 & 2.4 & 2.4 \\ 
   & 2 & 40 & 2.8 & 5.2 \\ 
   & 3 & 86 & 6.0 & 11.2 \\ 
   & 4 & 173 & 12.0 & 23.2 \\ 
   & 5 & 389 & 27.1 & 50.3 \\ 
   & 6 & 714 & 49.7 & 100.0 \\ 
   \hline
 & all & 1437 & 100.0 &  \\ 
   \hline
\hline
Knowledge & 1 & 141 & 9.8 & 9.8 \\ 
   & 2 & 301 & 20.9 & 30.8 \\ 
   & 3 & 286 & 19.9 & 50.7 \\ 
   & 4 & 257 & 17.9 & 68.5 \\ 
   & 5 & 374 & 26.0 & 94.6 \\ 
   & 6 & 78 & 5.4 & 100.0 \\ 
   \hline
 & all & 1437 & 100.0 &  \\ 
   \hline
\hline
Channel & 1 & 436 & 30.3 & 30.3 \\ 
   & 2 & 208 & 14.5 & 44.8 \\ 
   & 3 & 756 & 52.6 & 97.4 \\ 
   & 4 & 37 & 2.6 & 100.0 \\ 
   \hline
 & all & 1437 & 100.0 &  \\ 
   \hline
   \hline
Gender & f & 122 & 40.3 & 40.3 \\ 
   & m & 181 & 59.7 & 100.0 \\ 
   \hline
 & all & 303 & 100.0 &  \\ 
   \hline
\hline
Location &  Turin & 122 &  40 & 40 \\ 
   &  Milan & 12  & 3 & 43 \\ 
    &  Napoli & 12  & 3 & 46 \\ 
     &  Others  & 157  & 54  & 100.0 \\ 
   \hline
 & all & 303 & 100.0 &  \\ 
    \hline
     \hline
Management & Yes & 220 & 72.6 & 72.6 \\ 
   & No & 66 & 21.7 & 94.3 \\ 
   & NA & 17 & 5.6 & 100.0 \\ 
   \hline
 & all & 303 & 100.0 &  \\ 
   \hline
\hline

\end{tabular}
\endgroup
\label{tab: nominal}
\end{table}

\begin{table}[ht]
\vspace*{1 cm}
\centering
\caption{Descriptive statistics: continuous variables.} 
\begingroup\footnotesize
\begin{tabular}{lrrrrrrrr}
 \textbf{Variable} & $\mathbf{n}$ & \textbf{Min} & $\mathbf{q_1}$ & $\mathbf{\widetilde{x}}$ & $\mathbf{\bar{x}}$ & $\mathbf{q_3}$ & \textbf{Max} & $\mathbf{s}$  \\ 
  \hline
  Load & 303 &  1 &  7 & 15 & 20.1 & 26 & 202 & 21.1  \\ 
Age & 303 & 27 & 39 & 48 & 46.4 & 53 &  68 &  8.8 \\ 
 Seniority & 303 &  0 &  7 & 17 & 17.2 & 27 &  40 & 10.6  \\ 

  \end{tabular}
\endgroup
\label{tab: cont}
\end{table}

\par To be able to distinguish between different levels of performance we consider $Efficiency$ as ordinal with non-interval outcomes (with minimum satisfaction coded as 1 and maximum as 6) while all the other  variables are continuous ($Load$, $Age$, $Seniority$), categorical ($Knowledge$ and $Channel$) and dichotomous ($Gender$). Table \ref{tab: corr} reports the results of the correlation analysis. 
\begin{table}[H]
\vspace*{1 cm}
\caption{Correlation table.} 
\begin{tabular}{lllllcl}
\hline
 & Efficiency & Knowledge & Channel & Gender & Age & Seniority \\ \hline
Efficiency &  &  &  &  &  &  \\
Knowledge & $0.381$* &  &  &  &  &  \\
Channel & $0.130$* & $0.019$* &  &  &  &  \\
Gender & $0.021$ & $0.071$ & $0.118$ &  &  &  \\
Age & $-0.062$* & $-0.007$* & $0.093$* & $0.588$* &  &  \\
Seniority & $-0.078$* & $0.016$* & $0.028$* & $0.559$* & $0.806$* &  \\
Load & $-0.102$* & $0.036$* & $-0.298$** & $0.399$* & $0.127$* & $0.234$* \\ \hline
\tiny
\end{tabular}
\newline * significant level at 99\%, for the Pearson correlation for numeric variables, polyserial correlation for numeric and ordinal variables and polychoric correlation for ordinal variables.
\label{tab: corr}
\end{table}
\par The variable reflecting employees' performance does not show strong correlations with either of the predictors. $Efficiency$ is moderately correlated with $Knowledge$, suggesting that a more complex task may result in higher efficiency. The correlation between $Gender$ and $Efficiency$ is non significant, while $Load$, $Seniority$, and $Age$ variables are negatively correlated with $Efficiency$. This preliminary evidence might corroborate the hypothesis that the abundance of social capital may indeed lead to the decrease in efficiency of communication and suggests that higher organizational tenure and personal experience may not be always seen as a predictor of a positive performance.

\subsection{The model}
\par In order to test the impact of the various variables on the ordinal categorical variable $Efficiency$  we made use of standard ordinal logistic regression, which is widely applied as an effective method for modeling categorical outcomes with respect to its order as a function of both continuous and categorical predictors \citep{Harrell2015}. Specifically we tested the following model:

\begin{equation}\label{eq:1}
    \log (\frac{p_{j}^{c}(x)}{1-p_{j}^{c}(x)})=\alpha_{j}+{\beta}'x  \quad  \forall j=1,...,6
\end{equation}

\begin{itemize}
\item $p_{j}^{c}(x)=P(Y=<j|X=x)$ represents the logit of the cumulative distributions where $Y$ is the ordinal dependent variable measuring the efficiency of the response with value $j$ from 1 to 6.
\item $x=(x_{1},...,x_{k})$ is the matrix of the $k\_{th}$ independent covariates.
\item $\alpha_{1},...,\alpha_{r-1}$ and $\beta_{1},...,\beta_{k}$ the regression coefficients to be estimated.
\end{itemize}

\par This regression method relies on cumulative distributions and fits parameters for each association to estimate a general trend across the ordinal values of the dependent variable, retaining information on the rank ordering \citep{Agresti2002}. 
\par One of the most fundamental assumptions of the ordinal logistic regression is the proportional odds assumption, which requires that different categories of the variable have the same effect on the odds. Following \citet{Harrell2015}, a visual estimation of the linear predicted values of the given variables calculated with relaxed parallel slope assumption suggested that the variable $Knowledge$ violates the proportional odds assumption.  The Brant test \citep{Brant1990} shows that the effect of the variable is not constant across separate binary models, which supports the idea that the proportional odds assumption is violated. As a solution we group the six categories of the variable $Knowledge$ in two categories which summarize the possible answers in "Simple Knowledge" and "Complex Knowledge". The former includes knowledge transfers, which require simple instruction or some more complex explanation of simple instructions, while the latter groups categories which involve practical demonstration with exercise and experiments, analysis of the specific problem or even a step-by-step follow up and a verification of the results. Repeated testing with the binary version of the variable did not yield significant results \footnote{${\chi}^2$ $=$ $39.75$ $(p > .31)$}, indicating that  proportional odds assumptions is now satisfied.


\subsubsection{Identification strategy}
It is very difficult to elicit causal relations in network cross-sectional data since we do not observe the network formation. 
Indeed, the observed snapshot might depend on some characteristics of the individual, say talent, which remain unobserved by us but not by the agents in the network. In the specific case of this paper, co-workers might use this information for the the decision of contacting a specific person and forming the edge. 

Thus, an unobserved talent of the individual can at the same time explains both the performance and the network structure, which we use to compute the main independent variable \textit{Load}. This raises an important concern about endogeneity of the model due to simultaneity and unobserved heterogeneity. Since, in this context, experimental data are not available, the recommended solution \citep{carpenter2012social} is the use of a viable instrument, which correlates with \textit{Load}, but not with the error term in model. There are very few works which  manage to put forward sensible instruments for network analysis and they usually rely on lagged-values \citep{zaheer2009network}. We are not aware of any instrument that can be generally applied in a network context, and, as further important contribution of the paper we put forward an instrument that can be employed in many different areas of causal modeling with network data.

As instrument for the endogenous regressor \textit{Load}, we use a weighted centrality measure computed exactly as \textit{Load}, but based on the structure of a different network which does not depend on individual decisions to obtain information, and, thus,  satisfies the exclusion restriction. Specifically, we use a network in which the edges are the degrees of similarity among nodes with respect to the variables $Gender$, $Age$, $Seniority$. 

The logic behind this approach rests on the concept of homophily, which describes the empirical stylized fact that similar people along different socio-demographic characteristics are more likely to form edges with one another, independently on the purpose of the edge formation \citep{mcpherson2001, jackson2010social, beretta2018cultural}. The concept is very close to the idea of assortative matching in labour economics or assortative mating in biology, according to which some animals with a similar phenotype, or socio-economic characteristics for the human species, mate with a higher frequency than expected if the mating were random \citep{robinson2017genetic}.

We compute the homophily network of the individuals in the dataset based on the characteristic we can observe such as \textit{Gender}, \textit{Age}, and \textit{Seniority}. As expected, the instrument correlates with the variable \textit{Load}, since, homophily usually hold in very different types of network.

The econometric challenge of adopting an instrument variable approach when dealing with non-linear models has recently seen a lively debate comparing the two-stages predictor substitution (2SPS) with the two-stages residual inclusion (2SRI) approach \citep[][among others]{palmer2017correcting, basu2017comparing, terza2018two}. As in a standard two-stages least-square (2SLS) instrumental variables regression for linear models, both these two approaches relies on a first stage estimation in which the endogenous variables is regressed over the exogenous variables and the exogenous instrument. However, the 2SPS approach maintains the idea of using the predicted value of the first step regression in the second step, while the 2SRI uses in the main regression both the endogenous regressors and the residuals from the first stage estimation. \citet{terza2008} show that specifically for the Ordered Multinomial Logit that 2SRI with a linear auxiliary first-stage function is a consistent estimator. In the next section, we show and comment results for different models.

\subsection{Results}
\par The results of the model are reported in the Table \ref{maintable}, which shows coefficients, p-values and standard errors. We run a battery of models. The first column shows the results of the standard ordered logistic model, while columns 2 and 3 for the instrumented version of the same model with methodologies both 2SLS and 2SRI. In columns 5 and 6, shows the results for the  Binary Logistic Model 2SR1 and 2SLS Linear Probability model with $Efficiency$ as a binary variable, as discussed later.

\begin{figure}[H]
\caption{Odds-ratio for significant effects in IV-ordered-LOGIT-2SRI}
\label{fig:odds}
\centering
\includegraphics[width=0.8\textwidth]{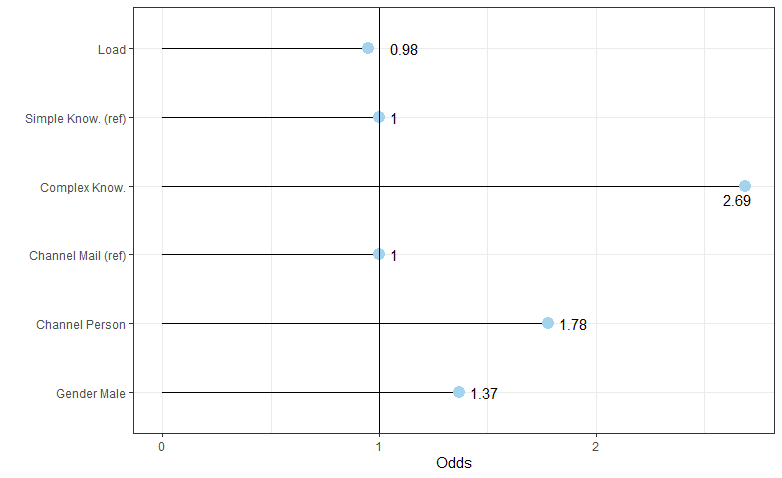}
\end{figure}

As described above, in the first three models, the dependent variable $Efficiency$ is expressed as categorical and ranging from lower to higher efficiency. Thus, a positive and significant value of the estimate indicates that when keeping all the independent variables constant, with one unit increase in a continuous covariate, the odds of moving to a higher level of efficiency increases. An easier interpretation of the result in figure \ref{fig:odds} is given by the odds-ratio. The odds ratio gives the factor of increase in the probability of moving to a higher category of efficiency when a continuous variable increases by one unit or a categorical variable change from the reference category. For instance, if the $Load$ increases by one unit only, the odds of moving to a lower category of $Efficiency$ increases by 1\%. The negative value of the impact of $Load$ upon $Efficiency$ corroborates the idea behind the paper. Since highly requested workers have a value for $Load$ about 200, it means that one additional request increases by 2\% the probability of moving in a lower category of $Perfomance$. 

Both the magnitude and the statistical significance of coefficients are stable across different models, with major differences for Linear Probability Model, which does not come unexpected since in that model the coefficient are marginal effects and not the marginal effect of the log-odds. 


\begin{table}[!htbp] \centering 
\small
  \caption{} 
  \label{maintable} 
\begin{tabular}{@{\extracolsep{5pt}}lccc|cc} 
\\[-1.8ex]\hline 
\hline \\[-1.8ex] 
 & \multicolumn{5}{c}{\textit{Dependent variable:}} \\ 
\cline{2-6} 
\\[-1.8ex] &\multicolumn{3}{c}{Efficiency} & \multicolumn{2}{c}{Binary Efficiency} \\ 
\cline{2-6} 
\\[-1.8ex] & \textit{Ordered}& \textit{ IV Ordered} & \textit{IV Ordered} & \textit{ IV Binary} & \textit{IV Linear} \\ 
 & \textit{Logistic} & \textit{Logistic 2SPS}& \textit{Logistic 2SRI} & \textit{Logistic 2SRI} & \textit{Probability Model}\\ 
\hline \\[-1.8ex] 
 Load / IV & $-$0.003$^{***}$ & $-$0.004$^{***}$ & $-$0.004$^{***}$ & $-$0.003$^{***}$ & $-$0.001$^{***}$ \\ 
  & (0.001) & (0.001) & (0.001) & (0.001) & (0.0003) \\ 

 Residual 1st-stage &  &  & 0.033$^{***}$ & 0.035$^{***}$ &  \\ 
  &  &  & (0.008) & (0.008) &  \\ 
 Knowledge & &&&& \\ 
 \hspace{0.5cm}Simple &  \multicolumn{5}{c}{\textit{Reference}} \\ 
      \hspace{0.5cm}Complex & 0.971$^{***}$ & 0.975$^{***}$ & 0.976$^{***}$ & 0.904$^{***}$ & 0.213$^{***}$ \\ 
  & (0.104) & (0.104) & (0.104) & (0.112) & (0.026) \\ 
   Channel & &&&& \\ 
   \hspace{0.5cm}Mail&  \multicolumn{5}{c}{\textit{Reference}} \\ 
 \hspace{0.5cm}Person & 0.461$^{***}$ & 0.437$^{***}$ & 0.436$^{***}$ & 0.467$^{***}$ & 0.109$^{***}$ \\ 
  & (0.123) & (0.123) & (0.124) & (0.134) & (0.031) \\ 
 \hspace{0.5cm}Phone & 0.019 & 0.021 & 0.015 & $-$0.038 & $-$0.009 \\ 
  & (0.156) & (0.156) & (0.157) & (0.177) & (0.041) \\ 
   \hspace{0.5cm}Other & $-$0.469 & $-$0.489$^{*}$ & $-$0.506$^{*}$ & $-$0.619 & $-$0.136 \\ 
  & (0.297) & (0.297) & (0.297) & (0.377) & (0.083) \\ 
  
 Seniority & $-$0.009 & $-$0.007 & $-$0.007 & $-$0.001 & $-$0.0003 \\ 
  & (0.010) & (0.010) & (0.010) & (0.010) & (0.002) \\ 
  
 Age & $-$0.011 & $-$0.012 & $-$0.012 & $-$0.015 & $-$0.003 \\ 
  & (0.012) & (0.012) & (0.012) & (0.013) & (0.003) \\ 
  Gender  & &&&& \\ 
   \hspace{0.5cm}Female &  \multicolumn{5}{c}{\textit{Reference}} \\ 
 \hspace{0.5cm}Male & 0.297$^{*}$ & 0.311$^{*}$ & 0.316$^{*}$ & 0.360$^{**}$ & 0.084$^{**}$ \\ 
  & (0.161) & (0.161) & (0.162) & (0.176) & (0.041) \\ 

 Constant &  &  &  & $-$0.025 & 0.488$^{***}$ \\ 
  &  &  &  & (0.465) & (0.108) \\ 
 
\hline \\[-1.8ex] 
Observations & 1,425 & 1,425 & 1,425 & 1,425 & 1,425 \\ 
R$^{2}$ &  &  &  &  & 0.072 \\ 
Adjusted R$^{2}$ &  &  &  &  & 0.067 \\ 
Log Likelihood &  &  &  & $-$925.620 &  \\ 
Akaike Inf. Crit. &  &  &  & 1,871.239 &  \\ 
Residual Std. Error &  &  &  &  & 0.483 (df = 1416) \\ 
Lipsitz test&  &&p-value = 0.053 &&\\
Pulkstenis-Robinson test&&&p-value = 0.560  && \\
Hosmer-Lemeshow test&&&p-value = 0.14  \footnotemark[6] && \\
\hline 
\hline \\[-1.8ex] 
\textit{Note:}  & \multicolumn{5}{r}{$^{*}$p$<$0.1; $^{**}$p$<$0.05; $^{***}$p$<$0.01} \\ 
\end{tabular} 
\end{table}

\begin{figure}[H] 
\caption{Predicted probabilities.}
\label{fig:lpm2}
\centering
\includegraphics[width=0.5\textwidth]{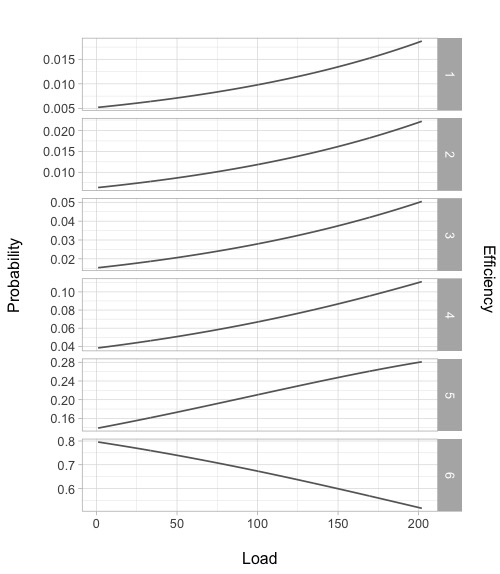}
\end{figure}

\begin{figure}[H]
\caption{Prediction for IV-LOGIT and IV-LPM }
\label{fig:lpm1}
\centering
\includegraphics[width=0.8\textwidth]{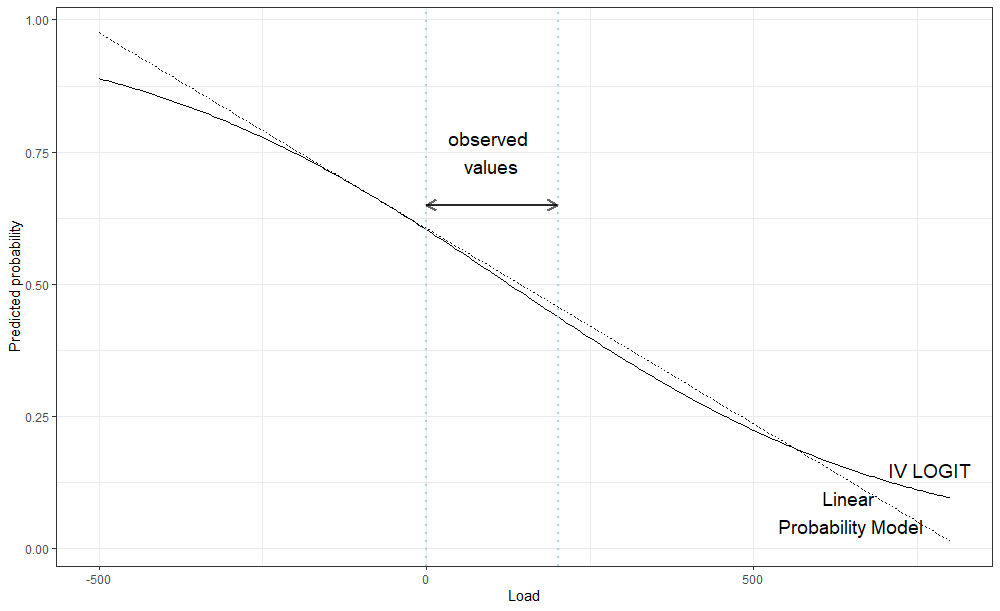}
\end{figure}

A different way to evaluate the impact is to look at the predicted probability of being in different category of $Efficiency$ for increasing value of $Load$ as in  Fig. \ref{fig:lpm2}. As expected for high level of $Load$ the predicted probability of high $Efficiency$ is much larger than for low values since employees contact colleagues who are likely to satisfy their request. However, it is also clear that with an increasing $Load$ the probability of scoring the highest level in $Efficiency$ dramatically drops and the probability is absorbed partly by all other categories. 

\par This evidence supports the idea  that the increased burden of collaborative behavior by employees tends to be associated with high level of efficiency, which eventually drops. 

\par As previously mentioned, the variable $Knowledge$ has been coded into a binary variable to satisfy OLR assumptions. Positive and significant estimates indicate that moving from simple type communication to more complex ones, such as "Step-by-step follow-up of the specific activity and verifying results" increases the odds of improving the $Efficiency$ by a factor of 2.74. The effect of the knowledge variable on the efficiency remains strong when controlling for the type of communication channel used. Data shows no significant effect on $Efficiency$ as the choice for the conversation shifts from email (reference category) to telephone. However, face-to-face collaboration had a significant positive impact on the dependent variable indicating that the odds of performing better are 1.71 higher.

\par Collaborative activity, knowledge complexity, and face-to-face interactions remain significant predictors of positive performance when personal characteristics are controlled for. The odds of an high evaluation of efficiency is 1.38 times larger when choosing a male over a female colleague. 

As Fig. \ref{fig:lpm2}, the large part of the negative effect of $Load$ on the performance reduce the $Efficiency$ from 6 to a lower value. This is an interesting result since highlight the burden information overload hits the most important nodes in the network of knowledge exchange. We exploit this result to performance an additional robustness check. We run a much simple and more robust models than the previous ones in which the variable $Efficiency$ can takes only two values, that either 6 or 0 for any other value. This simple model allows us also to report as benchmark the the Linear Probability Model, always in Table \ref{maintable}. Results \footnote{The outcome does not come with a surprise considering the graphical analysis from Fig. \ref{fig:lpm2} and the fact that that an ordered LOGIT generates similar output than a series of dichotomous one \citep{cole2004estimation}} are in line with the previous ones and Fig. \ref{fig:lpm1} which compares the predicted value for the two models using the binary efficiency in the observed support of $Load$, the IV-Logit and the IV-Linear Probability Model are overlapping. All in all, also the consistent IV-Linear Probability Model estimator provides similar results.

\subsection{Discussion}
\par The transition from hierarchical to matrix-based organizational structure, promotion of open-plan office space, and growing reliance on social collaboration tools facilitated the shift towards collaborative organizational culture. At the core of this shift is the higher visibility of employees intended to increase the frequency of interactions, promote creative thinking, and encourage innovation. The intensity of the information flows that employees experience daily under such working arrangements may also result in decreasing productivity of interactions. 
Based on the review of literature addressing possible adverse effects of extensive collaboration, we conclude that collaboration overload poses a risk to individual performance, contributes to work-related stress, burnout, and eventually can result in a higher attrition rate. 

\par The analysis of the communication practices in the organizational setting imposes several constraints, resulting in the lack of existing empirical research on collaboration overload. Compliance with data protection regulations limits the availability of data about informal networks of relationships.  We applied cross-sectional survey research to collect the data about working arrangements contributing to the increase of collaboration and estimate their impact on individual performance. Gathering information through survey methodology implies active participation of the subject under study and, as a result, raises significant concerns about endogeneity. 

\par Obtained results confirm that the creation of an informal network in the organization takes place independently of a formally codified hierarchical arrangement, and thus both types of networks must be considered for the accurate analysis of knowledge transfer processes. Neither time that an individual has spent in the organization, nor a higher level of management resulted as significant predictors of the rise in efficiency of expertise sharing. This suggests that to identify organizational knowledge repositories accurately, managers can not make an exclusive reference to the organizational structure, but have to also account for individual characteristics and existing informal networks.

\par Efficiency is positively associated with the nature of information complexity, employees gain high levels of satisfaction when they process requests that require more intense interactions, such as problem-solving or step-by-step follow-ups. Similarly, the probability of delivering higher level efficiency grows when the interaction happens in person rather than through email. Finally, neither organizational tenure nor the personal experience expressed in years is a significant factor, suggesting that the informal network is independent from the formal network of an organization's hierarchy. 

\par Undertaking research in the organizational setting of a private company poses a series of restrictions, privacy legislation being the most challenging for academic research. Limiting constraints on the data about employees' personal characteristics and a high cost of mining the information through observation and survey contribute to the poor quality of existing empirical research. 

\par We thus contribute to the research on information complexity and its requirements for higher cognitive capacity and time. Previous studies have suggested a link between the type of knowledge and the means and outcomes of its transfer from one actor to another \citep{ Schneider1987, Eppler2004}. This research provides sound empirical evidence that, with the rising complexity of knowledge, the intensity of interactions between actors will increase and thus the efficiency of collaboration will also increase. 

\par This analysis adds a new perspective on the efficiency of the methods for information transfer. There was no significant impact of the age of the actors, nor seniority. Considerable low levels of telephone usage compared to electronic mail and face-to-face conversation reveal employees' vision of an effective communication technology.  As expected, face-to-face communication delivers greater odds for providing higher productivity.

\par The results of the analysis suffer from a possible endogeneity because the formation of the observed networks depends very likely on the past and expected efficiency of the interaction. However, as discussed before, it is fair to assume that the negative effect of the collaborative load occurs later on in the network formation when it becomes excessive and creates time and cognitive constraints.  If results are biased, they are underestimating the negative effect of load on efficiency. 

\par  Managers are encouraged to incorporate the role of intraorganizational communications as a firm's strategic resource, with the relevant positive and negative externalities it may bring. Organizational decision-makers should embrace the importance of recognizing the regularities of social structures in order to understand the effect that a single actor behavior has on the welfare of other organizational players.
Managerial interventions should focus on two types of behavior change - directed on key actors and information seekers.
\par As the developed model suggests, the primary emphasis should be placed on identifying the key expertise holders among employees. Adequate coaching and mentoring programs facilitate the diffusion of knowledge possessed by stars. Quantifying collaborative load that organizational actors encounter in their everyday activity provides a benchmark for the identification of active collaborators who should be prioritized when developing retention and promotion plans. 

\par Development of knowledge repositories provides organizational leverage to control over the stream of expertise sharing requests.  Such organizational wikis summarise information that has a lower level of ambiguity and complexity,  thus can be easily codified and retrieved. Another intervention to equalize collaboration should be aimed at strategical staff placement. Organizational space arrangement should favor proximity of interdependent employees to allow for face-to-face collaborations, which increase the probability of delivering higher satisfaction of information request. 

\pagebreak
\section{Conclusions and Limitations}
\par This paper investigated the existence of collaborative overload in the workplace and gave an account of possible factors contributing to social capital failure. By modeling the social structure of collaborations in the professional service firm, we reproduce a network of individual incentives that translate into network outcomes. Consequently, we identify top performers that take upon themselves the majority of peer assistance requests and estimate the probability of an efficient outcome of the collaboration between two colleagues under specific task constraints.

\par We thus conclude that active involvement in collaborative activity can lead to over-embeddedness of top performers reducing the probability of the establishment of efficient knowledge sharing practices. In line with the previous research \citep{Eppler2004, Oldroyd2012, Cross2002, Abbasi2014}, the results of this study support the idea about the presence of few focal actors, who possesses greater visibility in the organizational network, acting as a reference point for the majority of colleagues. Secondly, with the growth of collaborative load, we observe a negative impact on the performance of those actors. The effect of collaborative load remains statistically significant even after controlling for communication mode, organizational context, and personal factors.

\par Finally, a number of potential limitations encountered during the design of this study should be mentioned. Despite various determinants used for the evaluation of social capital, there is a general consensus on the low quality of delivered results \citep{Falk1998} since it is very hard to empirically capture to concept of performance in communication. The data used for defining our variable of interest, was self-reported, and it is therefore a potential source of bias. While the survey methodology allows the type of communication appropriate for the research to be isolated, constructed network of organizational collaborations has a static nature, which makes causal inference extremely uncertain.  Moreover, this evidence is based on a specific case in the  professional services industry, limiting the external validity of the findings. Nevertheless, the data gathering has been very parsimonious and make it easy for future work the collection of similar data to replicate this exercise in a different context. 

\par All in all, for the first time we add a robust quantitative result to a growing body of qualitative literature on risks posed by the excess of information requests placed especially on a few key individuals, affecting their performance, and ultimately damaging the proper function of the organization. Results do not suggest the reduction of intraorganizational communications, but they rather signal  inefficiency in the knowledge and teamwork management. As managers stress the necessity and create more possibilities for teamwork, the pressure on key employees is rising. Application of this approach to organizational data enriches human resource departments with the knowledge of intra-organizational communication patterns and activity of knowledge workers, potentially at risk of being burdened by overload.

Failing in doing so  runs the risk of creating a disconnect in employees' engagement, thus contributing to the rise of barriers to inclusiveness. Though the value created by employees' interactions is technically hard to measure, it should not be seen as practically impossible.
\pagebreak

\clearpage
\section{Appendices}
\chapter{\textbf{Survey}} 
\label{survey}
\begin{enumerate}
\item \textbf{Indicate five colleagues to whom you came for help and from whom you received the necessary knowledge or information required to perform a certain task in the last six months?}
\begin{enumerate}
     \item Use the scale from 1 to 6 to evaluate how satisfying colleagues performance has been (1=minimum and 6 = maximum).
     \item Use the scale from 1 to 4 to indicate the frequency of interactions with the colleague (1=minimum and 4 = maximum). 
     \end{enumerate}

\item \textbf{Indicate the type of information that describes your request:}
\begin{enumerate}[label={\arabic*.}]
\item Simple instructions.
\item Instructions with more than one interaction for more clarifications.
\item More complex explanation of simple instructions, with greater interaction.
\item Practical demonstration with exercises and experiments on cases.
\item Analysis of the specific problem and definition of how to deal with it.
\item Step-by-step follow-up of the specific activity and verifying results.
\end{enumerate}
\item \textbf{Indicate the type of communication channel that was used for contacting a colleague:}
\begin{enumerate}[label={\arabic*.}]
\item Email
\item Telephone
\item Personal meeting
\item Other
\end{enumerate}
\end{enumerate}
\pagebreak

\clearpage
\bibliographystyle{abbrvnat}
\bibliography{reference}
\end{document}